\documentclass[twocolumn,preprintnumbers,amsmath,amssymb,prl]{revtex4-1}

\usepackage{graphicx}
\usepackage{dcolumn}
\usepackage{bm}
\usepackage{color}
\usepackage{ulem}

\begin{document}

\title{
Quasiparticle Nodal Plane in the Fulde-Ferrell-Larkin-Ovchinnikov State of FeSe 
}

\author{S.\,Kasahara$^{1, 4}$}
\author{H.\,Suzuki$^{1}$}
\author{T.\,Machida$^{2}$}
\author{Y.\,Sato$^{1}$}
\author{Y.\,Ukai$^{1}$}
\author{H.\,Murayama$^{1}$}
\author{S.\,Suetsugu$^{1}$}
\author{Y.\,Kasahara$^{1}$}
\author{T.\,Shibauchi$^3$}
\author{T.\,Hanaguri$^2$}
\author{Y.\,Matsuda$^1$}

\affiliation{$^1$Department of Physics, Kyoto University, Kyoto 606-8502, Japan}
\affiliation{$^2$RIKEN Center for Emergent Matter Science, Wako, Saitama 351-0198, Japan}
\affiliation{$^3$Department of Advanced Materials Science, University of Tokyo, Chiba 277-8561, Japan}
\affiliation{$^4$Research Institute for Interdisciplinary Science, Okayama University, Okayama 700-8530, Japan}



\begin{abstract}
The Fulde-Ferrell-Larkin-Ovchinnikov (FFLO) state, characterized by Cooper pairs condensed at finite momentum, has been a long-sought state that remains unresolved in many classes of fermionic systems, including superconductors and ultracold atoms. A fascinating aspect of the FFLO state is the emergence of periodic nodal planes in real space, but its observation is still lacking. Here we investigate the superconducting order parameter at high magnetic fields $\bm{H}$ applied perpendicular to the $ab$ plane in a high-purity single crystal of FeSe. The heat capacity and magnetic torque provide thermodynamic evidence for a distinct superconducting phase at the low-temperature/high-field corner of the phase diagram. Despite the bulk superconductivity, spectroscopic-imaging scanning tunneling microscopy performed on the same crystal demonstrates that the order parameter vanishes at the surface upon entering the high-field phase. These results provide the first demonstration of a pinned planar node perpendicular to $\bm{H}$, which is consistent with a putative FFLO state.
\end{abstract}

\maketitle

Exotic superconductivity with a nontrivial Cooper pairing state has been a central subject in condensed matter physics. In conventional Bardeen-Cooper-Schrieffer (BCS) superconductors, superconducting state consists of singlet pairs with zero center-of-mass momentum ({\boldmath $k$}$\uparrow$, {\boldmath $-k$}$\downarrow$). In contrast, the FFLO state is a novel pairing state, in which the pair formation occurs between Zeeman-split parts of the Fermi surface and a new pairing state ({\boldmath $k$}$\uparrow$, {\boldmath $-k+q$}$\downarrow$) with finite center-of-mass momentum ${\bm q}$ is realized. The appearance of the ${\bm q}$-vector implies translational symmetry breaking, in addition to $U(1)$-gauge symmetry breaking.  As a result, the order parameter $\Delta(\bm{r})$ of condensed fermion pairs undergoes one-dimensional (1D) spatial modulation, leading to normal conducting planes (nodal planes), in the superconducting state~\cite{Matsuda07, Wosnitza18, Kinnunen18}. 

In superconductors, the FFLO state can appear when an external magnetic field $\bm{H}$ is applied, but the formation of this state requires stringent conditions; a sufficiently large Maki parameter, a ratio of the orbital to the Pauli-paramagnetic limiting upper critical fields $\alpha_M\equiv \sqrt{2} H_{c2}^{orb}/H_{c2}^{Pauli}\approx 2(m^*/m_e)(\Delta_0/\varepsilon_F)>1.5$, is a prerequisite, where $m_e$ is the free electron mass, $m^*$ is the effective mass, $\Delta_0$ is the superconducting gap magnitude at zero field and $\varepsilon_F$ is the Fermi energy~\cite{Matsuda07}. 
Promising candidate systems with a large Maki parameter are layered superconductors in a parallel field~\cite{Wosnitza18,Agosta17} and heavy fermion compounds with large $m^*$~\cite{Matsuda07,Radovan03,Bianchi03,Shishido11}. The former includes several quasi-2D organic superconductors and the latter includes CeCoIn$_5$. However, although some features of the field-induced superconducting phase have been reported in the above systems, a key observation of the FFLO state, i.e. the emergence of the nodal planes in real space in the superconducting state is still lacking.

The layered iron-chalcogenide FeSe ($T_c\approx$ 9\,K) has recently stirred great interest as a platform to investigate various exotic superconducting states~\cite{Hsu08,Shibauchi20,Kreisel20}. FeSe exhibits a tetragonal to orthorhombic structural transition at $T_s \approx 90$\, K without magnetic order~\cite{Boehmer18}. Fermi surface consists of a hole pocket at the zone center and one or two electron pockets at the zone boundary~\cite{Shibauchi20,Coldea18,Yi19,SM}. A remarkable feature is the emergence of high-field superconducting phases for both ${\bm H} \parallel ab$~\cite{Kasahara20,Hardy20,Ok20} and ${\bm H} \parallel c$~\cite{Kasahara14}.  In particular, for ${\bm H} \parallel ab$, a distinct first-order phase transition deep inside the superconducting phase, which is revealed by a discontinuous jump of the thermal conductivity, has been reported~\cite{Kasahara20}.  

The high-field superconducting phases of FeSe have been discussed in terms of the FFLO state~\cite{Kasahara14,Kasahara20}. In FeSe, the Maki parameter of each band has been determined microscopically by using spectroscopic-imaging scanning tunneling microscopy (SI-STM)~\cite{Kasahara14}, quantum oscillations~\cite{Terashima14}, and angle-resolved photoemission spectroscopy~\cite{Watson15a} measurements. For the $c$-axis field, $\alpha_M\approx 2.5$ and 5 for hole and electron pocket, respectively~\cite{Kasahara14}. Owing to the extremely large $\Delta_0/\varepsilon_F$ and mass enhancement caused by the electron correlation effect~\cite{Shibauchi20,Kasahara14,Kasahara20}, Maki parameters in both bands are strikingly enhanced, demonstrating that FeSe has a strong Pauli pair breaking effect. The FFLO state is stabilized when a large area of the spin-up Fermi surface is connected to the spin-down surface by the ${\bm q}$-vector. For a 2D system with a cylindrical Fermi surface, the two Zeeman-split Fermi surfaces touch on a line by a shift of the ${\bm q}$-vector in the 2D plane. Thus, in the layered compounds, the FFLO state with $\bm{H}\parallel\bm{q}$ is more favorable under a parallel field than in a perpendicular field~\cite{Shimahara94}. Therefore, whether the FFLO state is formed for ${\bm H} \parallel c$ is a highly non-trivial issue~\cite{Song19,Shimahara21}.

Importantly, in FeSe, the large Maki parameter for ${\bm H} \parallel c$ still fulfills the requirements for the possible FFLO phase even in this field orientation, providing an excellent platform to tackle the FFLO physics.
Here, we performed heat capacity, magnetic torque, resistivity, and SI-STM measurements on a high-quality single crystal of FeSe for {\boldmath $H$}$\parallel c$. All measurements were performed on the same crystal~\cite{SM}. This is particularly important because $T_c$ and $H_{c2}$ slightly depend on the samples. To reduce the $T_c$ distribution caused by the inhomogeneity, we selected a tiny crystal ($\sim 30\,\mu$g) containing an extremely small amount of impurities. The heat capacity demonstrates a distinct superconducting phase at the low-$T$/high-$H$ corner of the phase diagram.  Despite bulk superconductivity, SI-STM reveals that the superconducting order parameter vanishes at the surface upon entering the high-field phase. The combination of these thermodynamic and spectroscopic results supports the presence of the FFLO state.

\begin{figure}[t!]
	\begin{center}
		\includegraphics[width=0.92\linewidth]{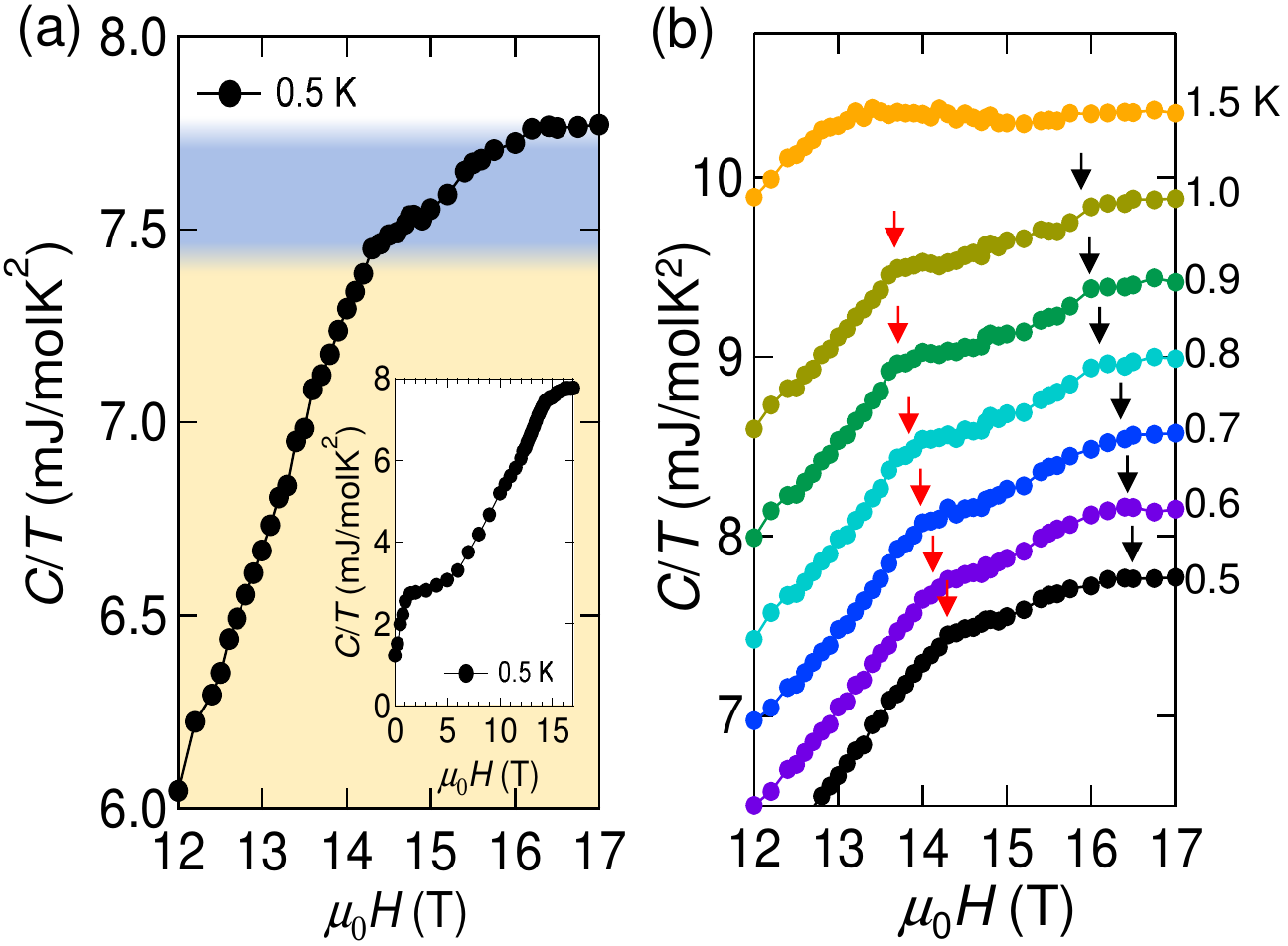}
		\caption{ (a) $H$-dependence of the heat capacity, $C(H)/T$, between 12 and 17\,T at 0.5\,K. The orange and blue shades denote the two different regimes of the $C(H)/T$ below and above $H^\ast$. The inset shows $C(H)/T$ for the whole $H$ range from 0 to 17\,T. 
		(b) $H$-dependence of $C(H)/T$ at different temperatures below $T = 2.0$\,K. Each curve is vertically shifted for clarity. The red arrows indicate the kink anomaly at $H^{\ast}$. The black arrows show the mean-field $H_{c2}^c$, above which the $C/T$ becomes $H$-independent. Two distinct kinks are observed up to 1\,K, but they are not seen clearly at 1.5\,K. }
	\label{fig:C}
	\end{center}
\end{figure}

As shown in the inset of Fig.\,\ref{fig:C}(a),  $C(H)/T$  grows steeply up to $\mu_0H \sim$1\,T and then increases gradually up to $\sim 6$\,T. Thus, a substantial portion of quasiparticles is already restored at $\sim$1\,T, which is consistent with multiband superconductivity~\cite{Sato18}. As depicted in the main panel of Fig.\,\ref{fig:C}(a),  $C(H)/T$ increases nearly linearly with $H$ above 12\,T. At $\mu_0H^*\approx 14$\,T, $C(H)/T$ exhibits a clear kink and then increases nearly linearly with a smaller slope above $H^\ast$.  At $\mu_0H\approx 16.4$\,T, $C(H)/T$ again exhibits a distinct kink and becomes $H$-independent at higher fields. As $C(H)/T$ is $H$-independent in the normal state, the kink at the higher field is attributed to the mean-field upper critical field $H_{c2}^c$. Thus, the present results provide thermodynamic evidence for the phase transition at $H^*$ well below $H_{c2}^c$. As shown by the blue shaded region in Fig.\,\ref{fig:C}(a), a substantial portion of the electrons remain condensed into a superconducting state just above $H^\ast$. We note that recent measurements of nuclear magnetic resonance (NMR) report the change of the spin-lattice relaxation rate $1/T_{1}T$ at $H^*$~\cite{Molatta21}. As shown by the red and black arrows in Fig.\,\ref{fig:C}(b), two distinct kinks are observable up to 1\,K. At 1.5\,K, however, they are not seen clearly, and $C(H)/T$ values are nearly $H$-independent above $\sim13$\,T. The nearly $H$-independent $C(H)/T$ at 1.5 K at high-$H$ is explained by  the strongly $T$-dependent specific heat jumps at $H_{c2}^c$, $\Delta C\propto T(dH_{c2}^c/dT)^2$. This jump structure is observable at high temperatures, but $\Delta C$ vanishes at very low temperatures. A sharp jump is often smeared out by thermal fluctuations (and inhomogeneity), and consequently, a broad peak appears just below $H_{c2}^c$. Such a peak structure has been reported above $\sim$2 K in FeSe \cite{Hardy20}. At 1.5\,K, although discernible peak structure is not observed, the $C(H)/T$ curve appears to be seriously influenced by the jump anomaly.

\begin{figure}[t]
	\begin{center}
		\includegraphics[width=0.76\linewidth]{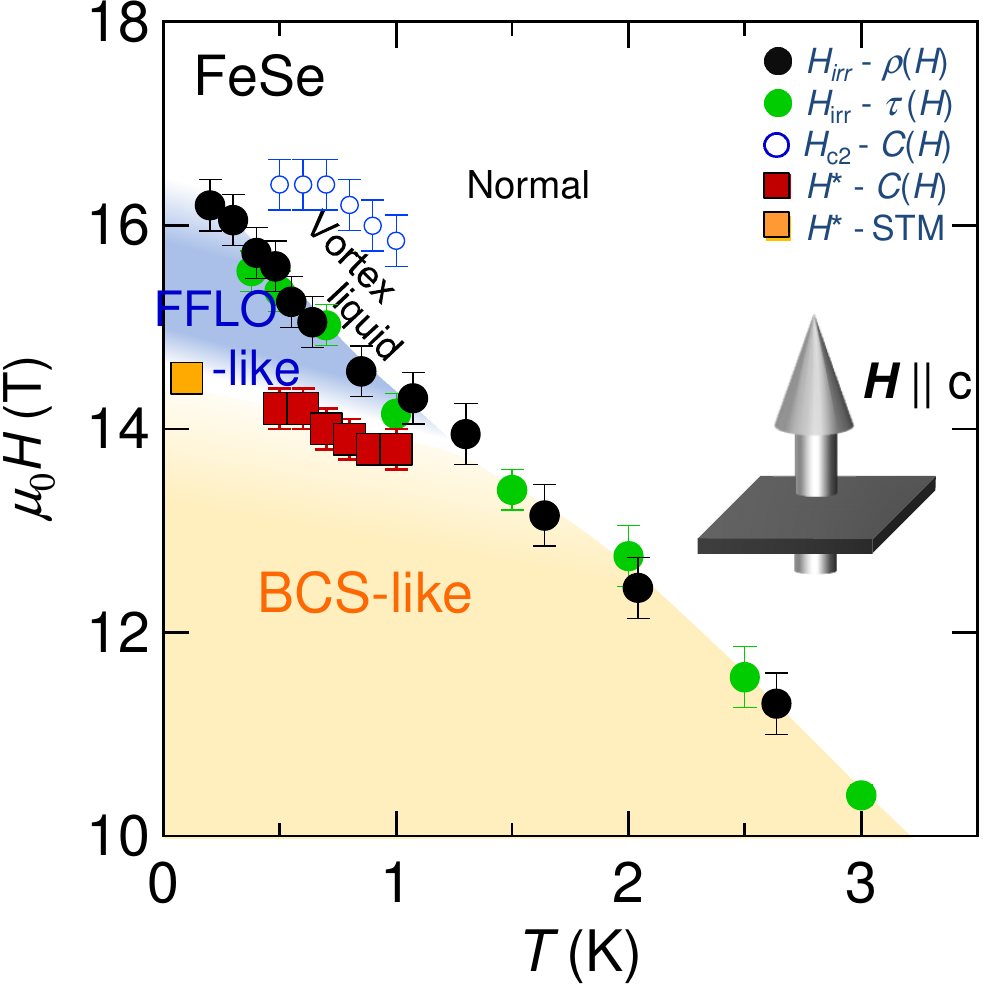}
		\caption {$H$-$T$ phase diagram of FeSe for ${\bm H} \parallel c$ at low temperatures. Blue open circles and red squares represent respectively the mean-field upper critical field $H_{c2}^c$ and $H^\ast$ that separates low- and high field superconducting phases, which are determined by the $C(H)/T$ curve. The orange square represents $H^{\ast}$ determined by SI-STM. Black and green circles show the irreversibility field $H_{irr}$ determined by resistivity and magnetic torque, respectively. The FFLO and BCS phases are indicated by blue and yellow shaded regions, respectively. }
	\label{fig:HTphase}
	\end{center}
\end{figure}

\begin{figure*}[t]
	\begin{center}
		\includegraphics[width=0.75\linewidth]{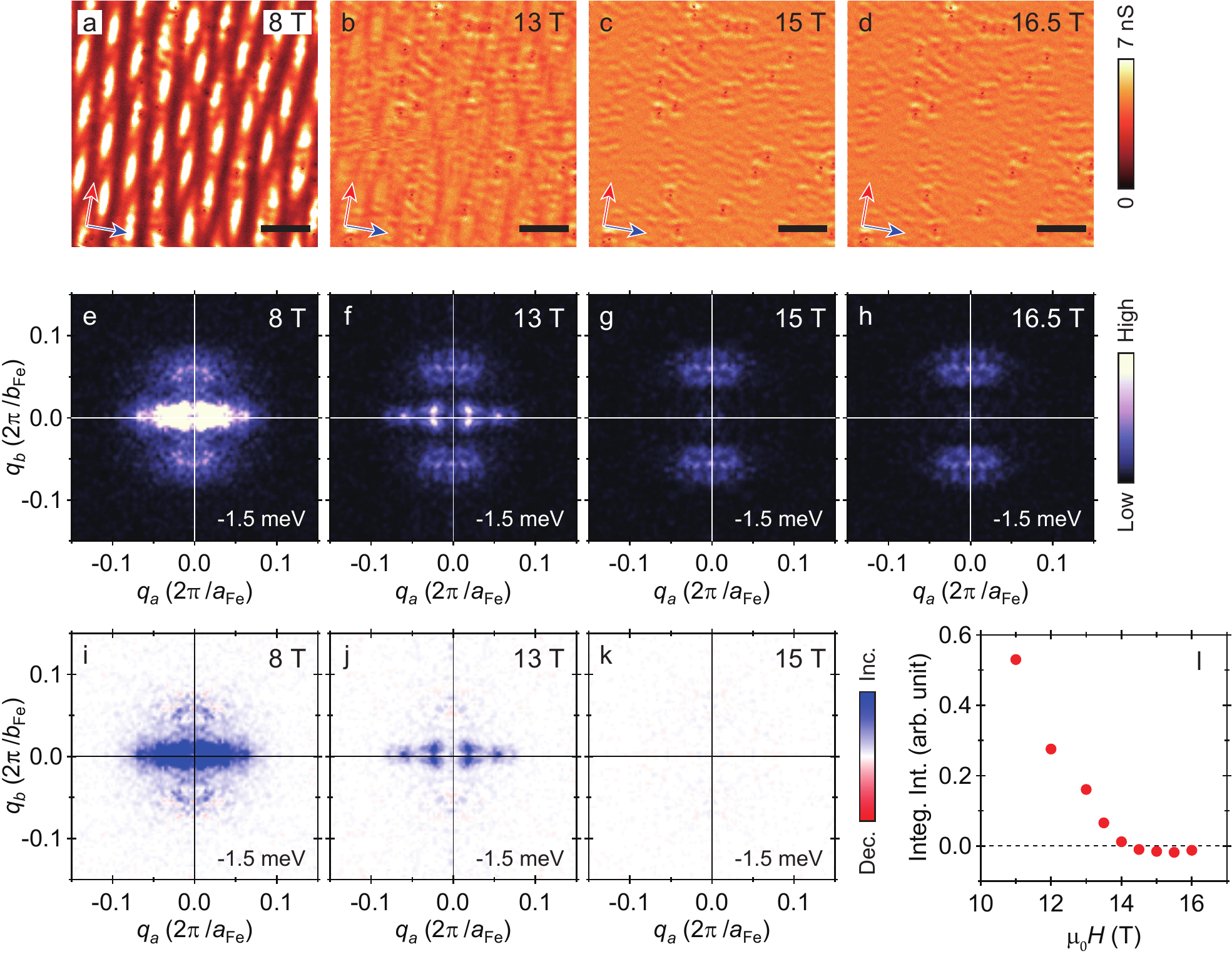}
		\caption{(a)-(d) $dI(E=0 {\rm\,meV},{\bm r})/dV$ images at various $H$ at $T\sim90$\,mK.
			Blue and red arrows denote ${\bm a}_{\rm Fe}$ and ${\bm b}_{\rm Fe}$, respectively, which are the primitive vectors of the Fe lattice. 
			We adopt the coordinate system $|{\bm a}_{\rm Fe}|<|{\bm b}_{\rm Fe}|$. Black scale 
			bars represent 20\,nm. 
			(e)-(h) Fourier-transformed $dI(E,{\bm r})/dV/(I(E,{\bm r})/V)$ images at -1.5\,meV. ${\bm q}_a$ and ${\bm q}_b$ are scattering vectors parallel to ${\bm a}_{\rm Fe}$ and ${\bm b}_{\rm Fe}$, respectively.
			(i)-(k) Difference images between Fourier-transformed $dI(E,{\bm r})/dV/(I(E,{\bm r})/V)$ below $H_{c2}^c$ and the one taken at 16.5\,T $>H_{c2}^c$.
			(l) $H$-dependence of the integrated intensity of the superconductivity-related signals. 
}
		\label{fig:STM}
	\end{center}
\end{figure*}

Next, we discuss the $H$--$T$ phase diagram. The irreversibility line separates a vortex-liquid from a vortex-solid region with zero resistivity. In the liquid regime, superconducting phase coherence is lost due to thermal or quantum fluctuations~\cite{Adachi03}. We determined the irreversibility field $H_{irr}$ by the $H$-dependence of the resistivity and the magnetic torque~\cite{SM}. Figure\,\ref{fig:HTphase} displays the $H$--$T$ phase diagram at high fields and low temperatures, showing $H^*$, $H_{c2}^c$, and $H_{irr}$. As shown in Fig.\,\ref{fig:HTphase}, $H_{irr}$ data determined by resistivity and torque coincide well with each other. It should be stressed that $H_{irr}$ locates between $H^*$ and $H_{c2}^c$. This firmly establishes that the long-range superconducting order survives above $H^*$. Thus the $H$--$T$ phase diagram displayed in Fig.\,\ref{fig:HTphase} indicates the presence of a noticeable high-field superconducting phase (blue shaded region), which is distinct from the low-field BCS phase (yellow shaded region). Although the precise determination of $H^\ast$ is difficult around 1.5 K, extrapolation suggests that $H^\ast$ merges with $H_{irr}$ around 1.5 K. As shown in Fig.\,\ref{fig:C}(a), the field slope of $C(H)/T$ above $H^*$ becomes less steep compared to that below $H^*$. This demonstrates that the upper critical field is enhanced by the formation of the high-field phase. Thus the present results provide thermodynamic evidence for the high-field superconducting phase.

To investigate how the superconductivity varies upon entering the high-field phase, we performed SI-STM at $\sim90$\,mK in a wide field range. Figures\,\ref{fig:STM}(a)-(d) display zero-energy tunneling conductance $dI(E,{\bm r})/dV$ images at various $H$, representing spatial distributions of the local density of states (LDOS) at the Fermi energy.
Here, $I(E,{\bm r})$ is the tunneling current at energy $E$ and position ${\bm r}$ and $V$ denotes the sample bias voltage. At 8\,T, vortices are imaged as high LDOS regions where the superconducting order is locally suppressed.
Each vortex is elliptical due to the nematicity~\cite{Song11,Watashige15,Hanaguri19} [Fig.\,\ref{fig:STM}(a)].
At 13\,T, although individual vortices are obscured, vortex-related LDOS inhomogeneity remains [Fig.\,\ref{fig:STM}(b)], meaning that superconductivity is still maintained. At 15\,T, the LDOS becomes uniform except for the quasiparticle interference patterns near defects [Fig.\,\ref{fig:STM}(c)], which are also seen in the normal state [Fig.\,\ref{fig:STM}(d)]. The absence of vortices at 15\,T indicates that the superconducting order diminishes at the surface even below bulk $H_{c2}^c$, presumably at $H^*$. To verify the relation between $H^*$ and the disappearance of superconductivity at the surface, we repeated the spectroscopic scan every 0.5\,T and quantified the vortex-related signals by separating them from the normal-state quasiparticle interference signals in Fourier space. Fourier-transformed spectroscopic images $dI(E,{\bm r})/dV/(I(E,{\bm r})/V)$ at $-1.5$\,meV are shown in Figs.\,\ref{fig:STM}(e)-(h). The vortex-related signals appear along ${\bm q}_a$ near $\bm{q}_b=0$, whereas the normal-state quasiparticle interference generates spots along ${\bm q}_b$~\cite{Hanaguri18}. The signals along ${\bm q}_a$ are confined below the superconducting-gap energy and are particle-hole symmetric~\cite{SM}, corroborating their superconducting origin.
We evaluated the vortex-related intensity by subtracting the Fourier-transformed $dI(E,{\bm r})/dV/(I(E,{\bm r})/V)$ in the normal state (16.5\,T) from the ones taken below $H_{c2}^c$ [Figures\,\ref{fig:STM}(i)-(k)], followed by the integration over $|E| < 6$\,meV. The resultant integrated intensity [Fig.\,\ref{fig:STM}(l)] vanishes at around 14.5\,T with a distinct kink. We conclude that this field corresponds to $H^*$ because it lies on an extrapolation of the $H^*$ from high temperatures (filled orange square in Fig.\,\ref{fig:HTphase}).

\begin{figure}[t]
	\begin{center}
		\includegraphics[width=0.75\linewidth]{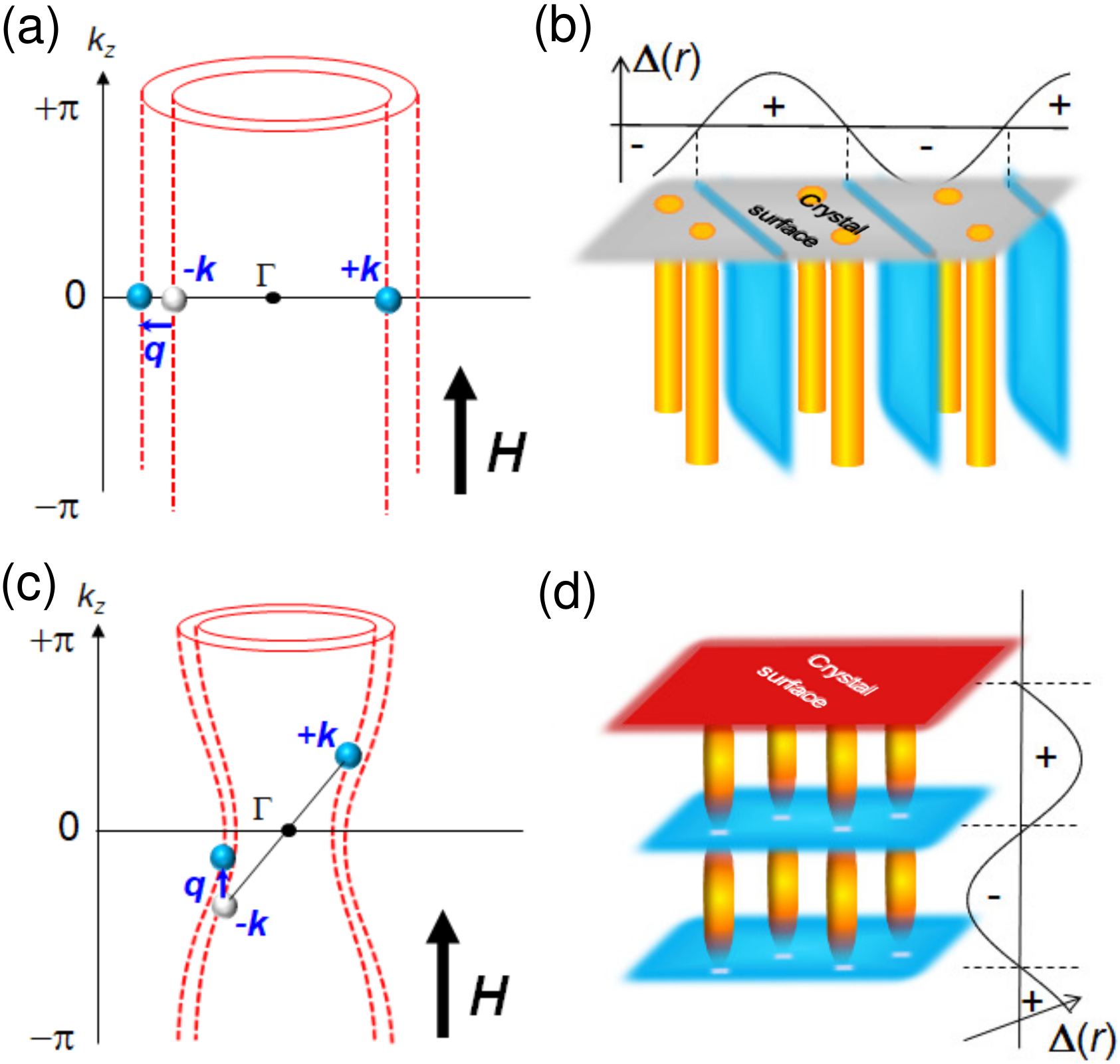}
		\caption{(a) Electron pairing with ({\boldmath $k$}$\uparrow$, {\boldmath $-k+q$}$\downarrow$) in a purely 2D Fermi surface. For the higher order FFLO state with $n > 0$, the ${\bm q}$ vector orients perpendicular to ${\bm H}$.  
(b) Schematic illustration of the superconducting order parameter $\Delta$ in real space for an $n > 0$ FFLO state. 
(c) Electron pairing with ({\boldmath $k$}$\uparrow$, {\boldmath $-k+q$}$\downarrow$) in a quasi-2D warped Fermi surface. The ${\bm q}$ vector orients along the direction of ${\bm H}$. (d) Schematic illustration of the superconducting order parameter $\Delta$ for the $n = 0$ FFLO state. The oscillations of the order parameter and the periodic planar nodes appear along the direction of ${\bm H}$. } 
\label{fig:pairing}
	\end{center}
\end{figure}

The most striking feature is that the superconducting signal detected by SI-STM vanishes above $H^*$ [Fig.\,\ref{fig:STM}(l)], whereas $C(H)/T$ indicates that $\sim 5$\,\% of electrons remain condensed in the superconducting state at $H^*$ [Fig.\,\ref{fig:C}(a)]. These contrasting observations cannot be accounted for by the multiband effect but can be naturally explained by the formation of the planar node at the top surface of the {\textit ab} plane. Indeed, this is an expected fingerprint of the FFLO state, which has been suggested to be realized in FeSe at high fields.We point out that the formation of some sort of magnetic or charge orders would not be the origin of the field-induced superconducting phase. The torque and NMR measurements~\cite{Molatta21} show no evidence of the magnetic order. The formation of charge order is unlikely because LDOS images above $H^*$ [Figs.\ref{fig:STM}(c), (d), (g), and (h)] show no discernible features other than the quasiparticle interference patterns associated with the normal-state band structure.

We note that the vanishing of the superconducting order parameter at the surface is highly unlikely to be caused by the crystal inhomogeneity. First, we confirmed that the superconducting-gap spectra are reproducible from cleave to cleave, which guarantees the homogeneity of our crystal~\cite{SM}. Second, any signature of degradation was not detected at the surface investigated in this work~\cite{SM}. Third, the heat capacity for a crystal (81\,$\mu$g) obtained from the same batch was reported to exhibit a very sharp superconducting transition~\cite{Mizukami21}. Since we used a smaller crystal ($\sim$30\,$\mu$g) in the present study, the inhomogeneity is expected to be even smaller.

There are two possible scenarios of the FFLO state for ${\bm H} \parallel c$, in which  ${\bm H}$ is applied parallel to the quasi-2D cylindrical Fermi surface. In the first scenario, the Fermi surface is assumed to be purely 2D. In this case, an FFLO state, in which the ${\bm q}$-vector in the $ab$ plane connects the Zeeman split Fermi surface sheets,  is stabilized [Fig.\,\ref{fig:pairing}(a)]. The order parameter oscillates perpendicular to ${\bm H}$, and periodic nodal planes appear parallel to the vortices [Fig.\,\ref{fig:pairing}(b)]. This state corresponds to the FFLO state involving Cooper pairs with higher-order Landau levels of index $n\ge 1$~\cite{Matsuda07,Shimahara94,Song19,Shimahara21}. However, we can exclude this scenario because the superconducting order parameter remains finite in some regions of the $ab$ plane, which is inconsistent with the uniform suppression of superconductivity at the top surface. In the second scenario, warping of the cylindrical Fermi surface along the $c$ axis is assumed to be important. In this scenario, the FFLO state of the lowest Landau level $(n=0)$, in which the ${\bm q}$-vector parallel to the $c$ axis connects the warped regime of the Zeeman split Fermi surface sheets, is stabilized [Fig.\,\ref{fig:pairing}(c)]~\cite{Shimahara21}. The order parameter oscillates along the $c$ axis as $\Delta(z)\propto \sin(qz)$, and the periodic planar nodes appear perpendicular to the vortices [Fig.\,\ref{fig:pairing}(d)]. To place the FFLO nodal planes, it is natural to consider that one nodal plane is pinned at the top surface of the $ab$ plane, which is energetically favored. This situation can account for the experimental observations. We note that as the vortices are heavily overlapped near $H^\ast$, inhomogeneity of the quasiparticle distribution is largely suppressed. Above $H^\ast$, the system becomes more homogeneous due to additional quasiparticle excitations caused by the planar node formation. This is consistent with the recent NMR results, which suggest a homogeneous electronic structure above $H^\ast$~\cite{Molatta21}.

Next, we consider the influence of nematicity on the FFLO formation. The Fermi surface of FeSe consists of hole cylinders around the zone center and compensating electron cylinders around the zone corner [Fig.\,S1(a)]. The superconducting gap function is highly anisotropic within the $ab$ plane, as shown by the elongated vortices [Fig.\,3(a)]. It has been reported that the gap function has nodes or deep minima along the long axis of the elliptical hole pocket [Fig.\,S1(b)]. It has been pointed out that the pairing interaction is strongly orbital dependent and the $d_{xz}$ orbital in the hole pocket plays an important role for the superconductivity~\cite{Sprau17}. Moreover, the $d_{xz}$ orbital has larger weights on the flattened parts of the Fermi surface than the other portion~\cite{SM}. Therefore it is likely that these parts of the Fermi surface with $d_{xz}$ character play an essential role for the FFLO state not only for ${\bm H}\parallel ab$~\cite{Kasahara20} but also ${\bm H} \parallel c$. 
The phase boundary between the BCS and FFLO states is nearly $T$-independent for ${\bm H} \parallel ab$, whereas it increases as the temperature is lowered for ${\bm H} \parallel c$.  This may be caused by the orbital pair-breaking effect, which is relatively more important for ${\bm H}\parallel c$.

In summary, we provide thermodynamic evidence for a distinct field-induced superconducting phase at the low-$T$/high-$H$ corner of the phase diagram in FeSe for ${\bm H} \parallel c$. Remarkably, despite bulk superconductivity, SI-STM reveals that the superconducting order parameter vanishes at the surface upon entering the high-field phase. These results imply that the planar node induced perpendicular to {\boldmath $H$} is pinned at the surface, which is consistent with those expected for the putative FFLO state. Emergent high-field phases for both field directions are perhaps the most prominent features in the BCS-BEC crossover superconductor of FeSe, which would provide an unprecedented platform of the long-sought FFLO state. \\

We thank H. Adachi, I. Eremin, M. Ichioka, R. Ikeda, K. Ishida, S. Kitagawa, H. K\"{u}hne, T. Mizushima, S. Molatta, Y. Nagai, H. Shimahara,  J. Wosnitza, and Y. Yanase for stimulating discussion. This work is supported by Grants-in-Aid for Scientific Research (KAKENHI) (Nos. 15H02106, 15H03688, 15KK0160, 18H01177, 18H05227, 19H01855, 19H00649, 21H04443) and on Innovative Areas ``Quantum Liquid Crystals" (No. 19H05824) from the Japan Society for the Promotion of Science, and JST CREST (JPMJCR19T5). \\

\eject
\
\newpage
\onecolumngrid

\begin{center}
{\large \bf Supplemental Material: Quasiparticle Nodal Plane  in the Fulde-Ferrell-Larkin-Ovchinnikov state of FeSe }

\bigskip

{S.\,Kasahara$^{1, 4}$}, 
{H.\,Suzuki$^{1}$}, 
{T.\,Machida$^{2}$}, 
{Y.\,Sato$^{1}$}, 
{Y.\,Ukai$^{1}$}, 
{H.\,Murayama$^{1}$}, 
{S.\,Suetsugu$^{1}$}, 
{Y.\,Kasahara$^{1}$}, 
{T.\,Shibauchi$^3$}, 
{T.\,Hanaguri$^2$}, 
{Y.\,Matsuda$^1$}

\medskip
{\it \small
\setlength{\baselineskip}{16pt}
$^1$ Department of Physics, Kyoto University, Kyoto 606-8502 Japan\\
$^2$RIKEN Center for Emergent Matter Science, Wako, Saitama 351-0198, Japan\\
$^3$Department of Advanced Materials Science, University of Tokyo, Chiba 277-8561, Japan\\
$^4$Research Institute for Interdisciplinary Science, Okayama University, Okayama 700-8530, Japan
}
\end{center}

\section{Fermi surface and orbital dependent superconducting gap}
In FeSe, it has been reported that the Fermi pockets in the nematic state are extremely small and shallow. Figure\,S1(a) shows a schematic illustration of the Fermi surface of FeSe in the nematic state in the unfolded Brillouin zone proposed in Ref.~[15]. The Fermi surface consists of a hole pocket at the zone center and an electron pocket at the zone corner. Green, red, and blue areas represent the Fermi-surface regions dominated by $d_{yz}$, $d_{xz}$ and $d_{xy}$ orbital characters, respectively. Note that we use the coordinate, in which the two-dimensional Fe lattice is used as a principal unit cell and the nearest neighbor Fe-Fe distance is larger along the $b_{\rm Fe}$ axis ($y$ direction) than along the $a_{\rm Fe}$ axis ($x$ direction). The Fermi energies $\varepsilon_F^{h(e)}$ of the hole (electron) pockets are extraordinarily small, $\varepsilon_F^{h} \sim 10$ - 15\,meV and $\varepsilon_F^{e} \sim$ 5 - 10\,meV. 
In Fig.\,S1(b), we show a schematic figure on the amplitude of the superconducting gap $\Delta$ at the hole pocket. It has been shown that $\Delta$ is highly anisotropic in FeSe and the largest superconducting gap is observed for the flat portion with dominant $d_{xz}$ character.

\begin{figure}[h]
	\includegraphics[width=0.485\linewidth]{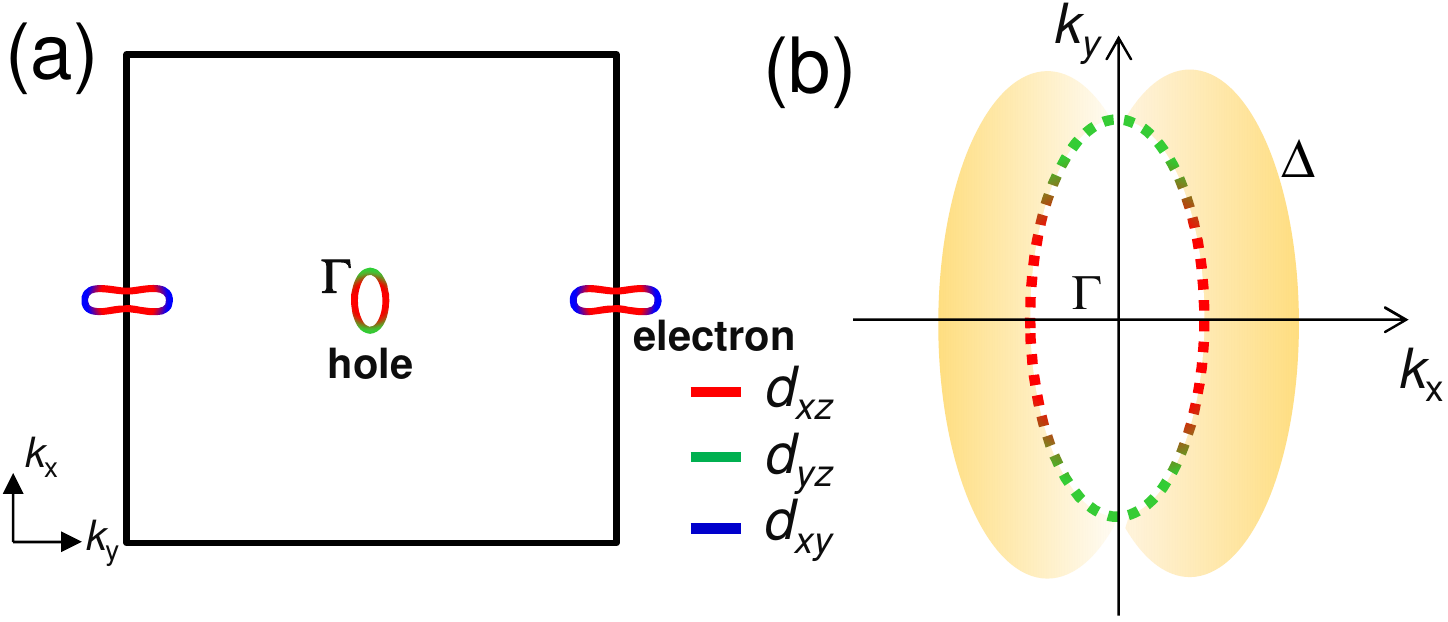}
\end{figure}

\noindent
{Figure S1. (a) Schematic figure of the Fermi surface of FeSe in the nematic state.  (b) In-plane anisotropy of the superconducting gap amplitude $\Delta$ at the hole pocket. The orange shade represents the amplitude of $\Delta$.}

\section{Sample characterization}
Single crystals of FeSe were grown by the vapor transport method. 
Particular attention was paid to select the crystal with extraordinary high quality. By measuring the STM topography, we confirmed that the investigated surface is atomically clean with only 2-3 defects per 10,000 Fe atoms [Fig.\,S2(a)]. This same single crystal was further cleaved into two pieces. One (smaller) piece ($\sim 250 \times 100 \times 5$\,$\mu$m$^3$) was used for the resistivity measurements by directly soldering the four indium contacts. The other (larger) piece  ($\sim$\,30\,$\mu$g) was used for the magnetic torque and the heat capacity measurements. In Fig.~S2(b), we show the temperature dependence of resistivity. The crystal exhibits zero resistivity at $T_c = 9.0$\,K. In the inset of Fig.\,~S2(b), we plot the temperature dependence of the magnetization $M$ measured by a SQUID magnetometer.  A sharp superconducting transition in the $M$-$T$ curve demonstrates that the sample is homogeneous. In Fig.~S2(c), we show zero-field superconducting-gap spectra taken on several FeSe surfaces obtained by different cleaves. The samples are all taken from the same batch. Spectra are averaged over the field of view wide enough to flatten the short-wavelength inhomogeneity due to the quasiparticle interference. Even fine structures of the spectra are quantitatively reproduced. Small variations among the spectra may be due to the difference in the exact nature of the tip used in different runs. Such cleave-to-cleave reproducibility guarantees that our samples are uniform and diminishing superconductivity at $H^* < H_{c2}$ is not due to the sample inhomogeneity.

\newpage

\begin{figure}[h]
	\includegraphics[width=0.80\linewidth]{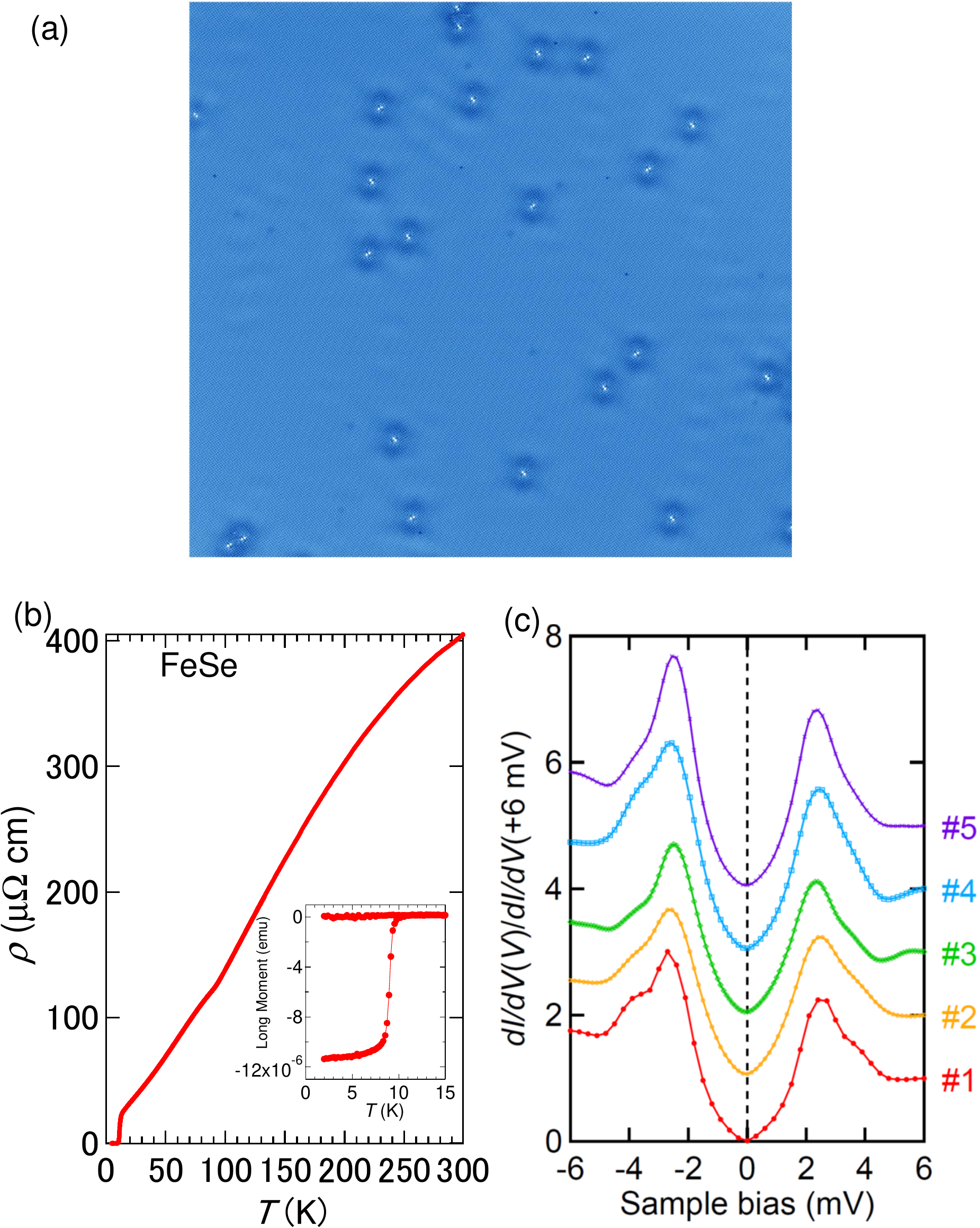}
\end{figure}

\noindent
{Figure S2. Sample characterization of FeSe.} (a) Constant-current STM topographic image of FeSe obtained at 90\,mK taken over the same field of view (FOV: $100 \times 100$ nm$^2$) of the SI-STM images shown in Figs\,3(a)-(d). Feedback conditions are set-point current $I=100$\,pA and set-point bias voltage $V=+20$\,mV. (b) Temperature dependence of resistivity measured on the same single-crystalline sample used for STM. The inset shows the temperature dependence of the magnetization under field-cooling and zero-field-cooling conditions in a magnetic field of 1\,Oe applied along the $c$-axis. (c) Zero-field superconducting-gap spectra averaged over the FOVs, which were taken on several FeSe surfaces obtained by different cleaves. Each spectrum is normalized at the value at +6\,meV and shifted vertically for clarity. Measurement conditions are summarized in Table S1. The spectrum \#1 has been measured for the same FOV investigated in this work.

\section{Torque magnetometry}
Magnetic torque $\tau$ was measured by the piezo-resistive micro-cantilever technique down to 0.38\,K and up to $\sim16$\,T. A tiny single crystal was carefully mounted onto the tip-less piezo-resistive lever (PRS-L450-F30-TL-STD, SCL-Sensor. Tech.) which forms an electrical bridge circuit. The field is slightly tilted away from the $c$ axis. 
In Fig.\,S3(a) and (b), we show the field dependences of $\rho$ and $\tau$, respectively. We determined $H_{irr}$ by the onset field of nonzero resistivity with a 0.5\,n$\Omega$\,cm criterion, and by closing field of hysteresis loops of magnetic torque to the level of 0.3\% of the whole signal.

\begin{figure}[h]
	\begin{center}
		\includegraphics[width=0.35\linewidth]{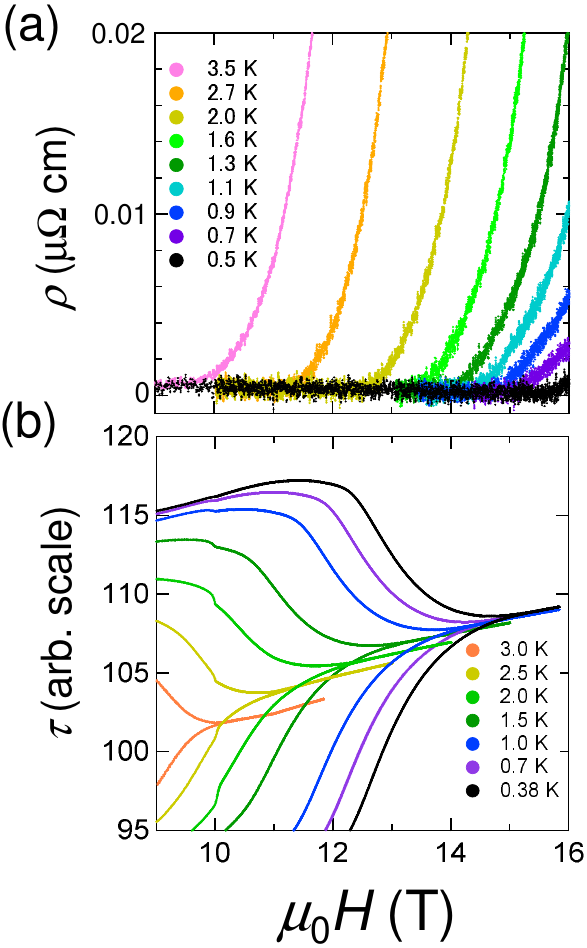}
	\label{fig:Hirr}
	\end{center}
\end{figure}
\noindent 
Figure S3. (a) Magnetic-field dependence of the resistivity $\rho$. (b) Magnetic-field dependence of the magnetic torque $\tau$.

\section{Heat capacity}
The heat capacity of the tiny single crystal of FeSe used for the STM and torque measurements was measured by the long-relaxation method \cite{Wang01,Taylor07}. 
With a tiny amount of grease, the sample was mounted onto the bare chip Cernox sensor, which is used as a thermometer and a heater. The sensor is suspended from the cold stage by gold-coated glass fibers such that it is weakly linked to the cold stage. The heat capacity of the crystal is obtained by subtracting the addenda from the total heat capacity measured with the sample. 

\section{SI-STM}
SI-STM experiments were performed with an ultrahigh vacuum dilution-fridge-based STM equipped with a 17.5\,T superconducting magnet~\cite{MachidaRSI}. We used a tungsten tip prepared by electrochemical etching. The tip was cleaned by field evaporation using a field-iron microscope, followed by controlled indentation into a clean Au(100) surface. The clean sample surface was obtained by vacuum cleaving at liquid nitrogen temperature. All data were taken in the constant-current mode with the feedback conditions of $I = 100$\,pA and $V = 20$\,mV. $dI/dV$ spectra were taken by standard lock-in technique with a bias modulation of 0.21\,mV$_{\rm rms}$. Whenever we changed the magnetic field, the sample was heated up above $T_c$ to ensure uniform vortex distribution in the sample.

\begin{table}
\caption{Measurement conditions for the spectra shown in Fig. S2(c). $V_{\rm mod}$ denotes bias modulation amplitude. 
}
	\begin{tabular}{ccccccc}
		\hline
Spectrum & FOV size (nm$^2$) & $I$ (pA) & $V$ (mV) & $V_{\rm mod}$ (mV$_{\rm rms}$)& $T$ (K) \\
		\hline
		\#1 & $100 \times 100$ & 100 & 20 & 0.21 & 0.09 \\
		\#2 & $160 \times 160$ & 100 & 20 & 0.21 & 1.5 \\
		\#3 & $50 \times 50$ & 100 & 10 & 0.11 & 1.5 \\
		\#4 & $100 \times 100$ & 100 & 20 & 0.21 & 1.5 \\
		\#5 & $160 \times 160$ & 100 & 20 & 0.21 & 1.5 \\
		\hline
	\end{tabular}
\end{table}

\begin{figure}[h]
 \begin{center}
	\includegraphics[width=\linewidth]{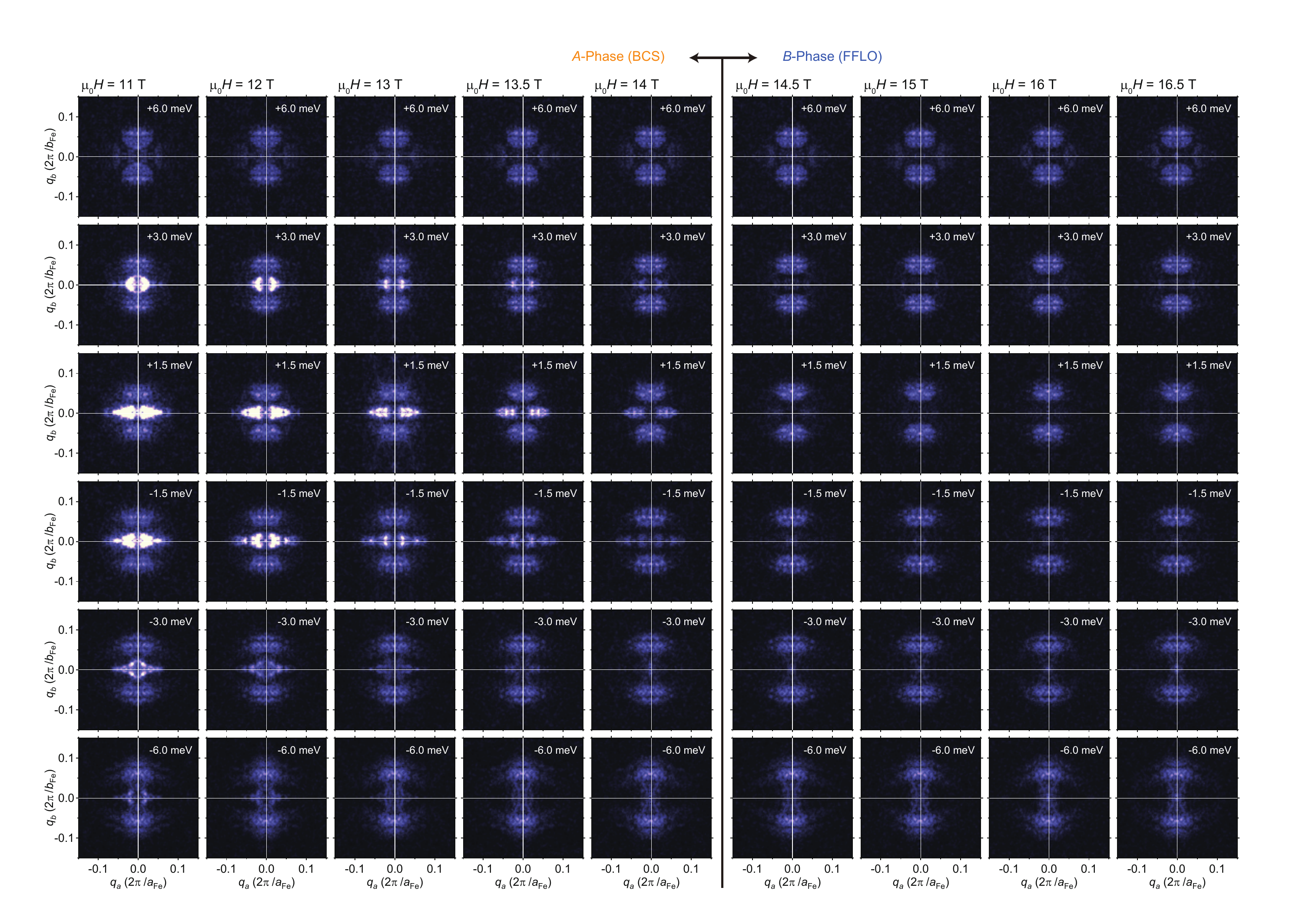}
		\end{center}
\end{figure}
\vspace{-5pt}
\noindent 
{Figure S4. Fourier-transformed spectroscopic images at 90\,mK.}
Complete dataset of Fourier-transformed spectroscopic images $dI(E,{\bm r})/dV/(I(E,{\bm r})/V)$ across $H^*$. Normal-state quasiparticle interference signals appear along ${\bm q}_b$, whereas density-of-states modulations associated with vortices are observed along ${\bm q}_a$. Note that we adopt the coordinate system $|{\bm a}_{\rm Fe}|<|{\bm b}_{\rm Fe}|$.

\newpage
\begin{figure}[h]
\begin{center}
	\includegraphics[width=\linewidth]{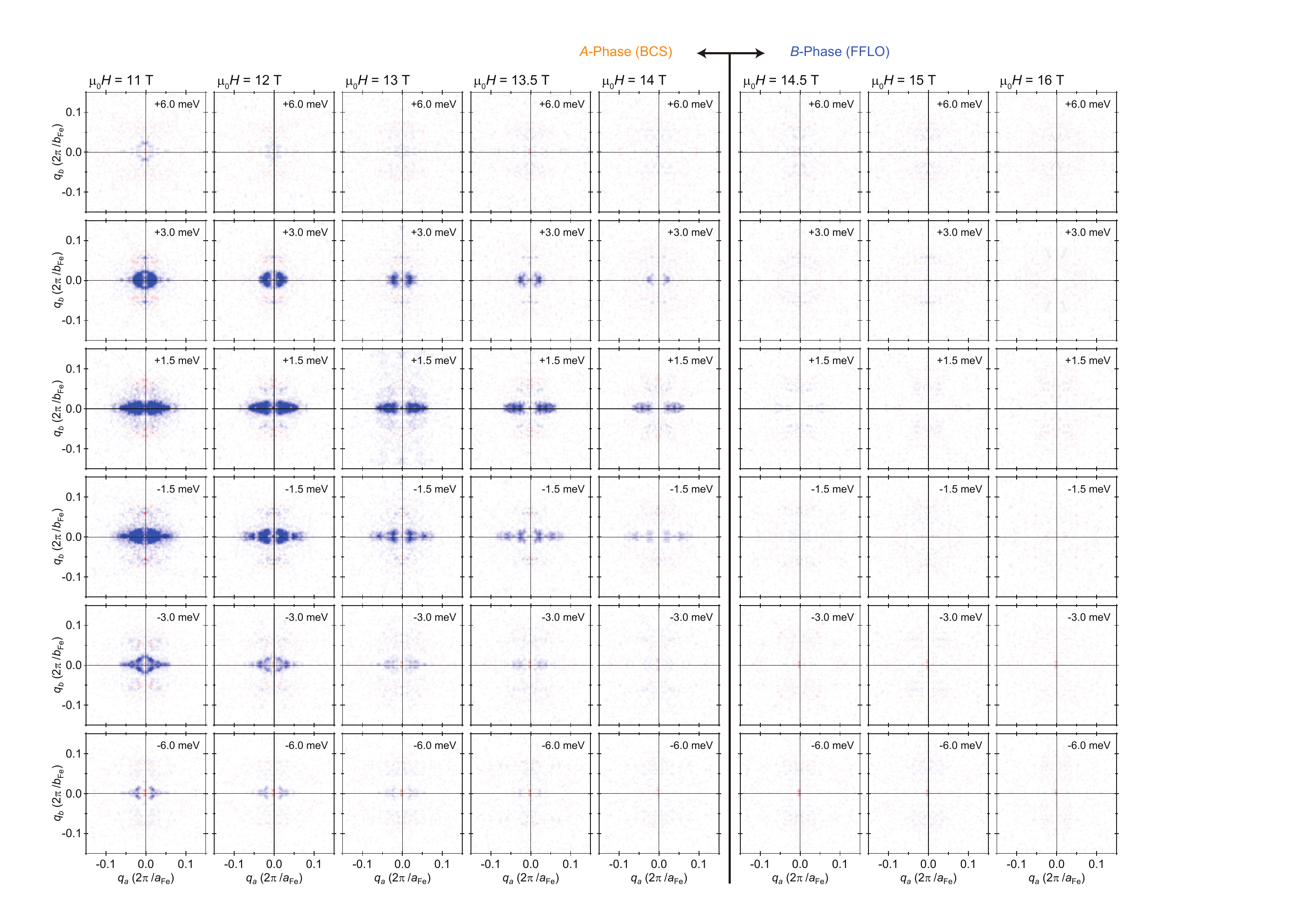}
\end{center}
\end{figure}

\noindent
{Figure S5. Superconducting signals in ${\bm q}$ space.} 
Superconducting signals obtained by subtracting $dI(E,{\bm r})/dV/(I(E,{\bm r})/V)$ at 16.5\,T $>H_{c2}^c$ from the ones taken under $H<H_{c2}^c$. The signals are confined below the superconducting-gap energy and disappear above $H^*$.

\end{document}